%Paper: hep-th/9209120
%From: ogievets@theor.jinrc.dubna.su ( Victor Ogievetsky)
%Date: Tue, 29 Sep 92 13:56:49 SET
%Date (revised): Sun, 18 Oct 92 20:38:42 +0300 (MSD)

\magnification=1200
\overfullrule=0pt
\parskip 3 pt plus 1pt minus 1 pt
\def\der#1{{\partial \over \partial #1}}
\font\sqi=cmssq8
\def\R{\rm I\kern-1.45pt\rm R}
\def\C{\kern2pt {\hbox{\sqi I}}\kern-4.2pt\rm C}

\line{\hfill CERN-TH.6653/92}
\vskip 2cm

\centerline{\bf SUPER SELF-DUALITY AS ANALYTICITY IN HARMONIC
SUPERSPACE}
\vskip 1cm
\centerline{Ch.Devchand }
\vskip 0.3cm
\centerline{Laboratory of Theoretical Physics,}
\centerline{JINR, Dubna, 141980, Russia}
\vskip 0.5cm
\centerline{and}
\vskip 0.5cm
\centerline{V. Ogievetsky
\footnote {$^*$} {On leave from the Laboratory of Theoretical Physics,
JINR, Dubna, 141980, Russia}}
\vskip 0.3cm
\centerline{Theory Division, CERN, }
\centerline{CH-1211 Geneva 23, Switzerland.}
\vskip 2cm
\centerline{\bf Abstract}
\vskip 0.5cm
A twistor correspondence for the self-duality equations for
supersymmetric Yang-Mills theories is developed. Their solutions
are shown to be encoded in analytic harmonic superfields
satisfying appropriate generalised Cauchy-Riemann conditions.
An action principle yielding these conditions is presented.

\vskip 1cm
 CERN-TH.6653/92

September 1992
\footline={\hfil}
\vfill\eject
\pageno=1
{\bf 1.} There has recently been a revival of interest in self-duality
equations, arising from numerous confirmations of a remarkable
suggestion [1] that all integrable systems are obtainable by dimensional
reduction from 4D self-dual theories. The purpose of this letter is to
show that for
supersymmetric Yang-Mills theories [2] the self-duality equations can
be written in a form in which their integrability becomes manifest
and their solutions can be constructed in terms of superfields
which are in a definite sense holomorphic. In other words, we shall
establish the  so-called ``twistor correspondence'' for the supersymmetric
self-duality equations analogous to that for the
ordinary ($N = 0$) self-duality equations [3], which in the harmonic
space language [4-6] involves a splitting of the coordinate
$x^{\alpha {\dot \alpha}}$ into
$x^{+\alpha} = u^+_{\dot \beta} x^{\alpha {\dot \beta}}$ and
$x^{-\alpha} = u^-_{\dot \beta} x^{\alpha {\dot \beta}}$,
where $u^\pm_{\dot \beta}$ are harmonics on the
two-sphere, which appears in the harmonisation of the rotation group
[4-6], and $\alpha$ and ${\dot \beta}$ are two-spinor indices.
The gist of ordinary self-duality is the Cauchy-Riemann-like equation
$$ \nabla^+_\alpha \phi = 0, \eqno(1)$$
where $\nabla^+_\alpha$ is the covariant derivative in $x^{- \alpha}.$
We shall show that this construction can be extended
naturally to supersymmetric gauge theories. In the
$N=1$ case we shall use the harmonic superspace with coordinates
$$ x^{\pm_\alpha} \equiv u^\pm_{\dot \beta} x^{\alpha{\dot \beta}},\
 \vartheta^\alpha,\ {\bar \vartheta}^\pm \equiv
 u^\pm_{\dot \alpha} {\bar \vartheta}^{{\dot \alpha}} ,\
  u^\pm_{\dot \alpha}.\eqno(2)$$
Now super-self-duality is the condition for the integrability of the
equation
$$ \bar{\cal D}^{+} \phi = 0,\eqno(3) $$
where $\bar{\cal D}^+$ is the gauge-covariant spinorial derivative
with respect to the variable $ \bar\vartheta^- $. $N = 1$
supersymmetric gauge theories respect chirality, so we may, without loss
of generality, take the field
$\phi$ in (3) to be  a {\bf chiral} superfield
$\phi(x^{+\alpha}, x^{-\alpha}, \bar \vartheta^+, \bar \vartheta^-)$
independent of $\vartheta^\alpha$, i.e.
$$ D_\alpha \phi = 0, \eqno(4)$$
where $D_\alpha$ is the covariant spinorial derivative
with respect to the variables $ \vartheta^\alpha,$ which may be made flat
by definition; and we shall set, throughout this letter, the
corresponding connections to zero, $$ A_\alpha = 0.\eqno(5)$$
Of importance is the fact that consistency of (3) and (4) implies
(1) in virtue of the algebra of spinorial derivatives, so
$N = 1$ self-duality implies the usual $N = 0$ self-duality.
We shall demonstrate that all solutions of the supersymmetric
self-duality conditions are encoded in a ``holomorphic'' chiral
superfield which satisfies generalised Cauchy-Riemann conditions;
in other words, we shall establish the famous twistor correspondence [3]
for the supersymmetric self-duality equations. By holomorphicity we mean
that there is a basis in which this superfield is independent
of $x^{-\alpha}, {\bar \vartheta}^-$ and $  \vartheta^\alpha.$
The resulting formulation greatly
simplifies the problem of constructing gauge superpotentials
$A_{\alpha{\dot \beta}}, A_{{\dot \beta}}$
solving the super-self-duality equations.
It also helps in the search for an action principle for super-self-duality.

Although our considerations are rigorous only for $4D$ Euclidean space,
we shall remain in the complexified picture with Lorentz group
$SL(2,\C)_L \times SL(2,\C)_R$, with $\alpha$ and ${\dot \alpha}$
labelling fundamental representations of $SL(2,\C)_L $ and $SL(2,\C)_R$,
respectively. In [6] it was shown that consideration of the complexified
picture is required by conformal invariance of the self-duality equations.
Reality conditions appropriate for the required signature of
four dimensional real space may be imposed; e.g. by identifying undotted
and dotted spinors as representations of the two different $SU(2)'$s in the
4D Lorentz group  $SU(2)_L \times SU(2)_R$ corresponding to a Euclidean
signature; or by identifying them as representations of two different
SL(2,\R)'s, with Lorentz group  $SL(2,\R)_L \times SL(2,\R)_R$
corresponding to a (2,2) signature. In the latter case we expect
intriguing peculiarities due to the non-compactness of $SL(2,\R)$
and an appropriate harmonic space needs to be considered (see e.g. [7]).

{\bf 2.} In complexified superspace ${\cal M}_{4|4}$ of complex dimension
$(4| 4)$ with coordinates $(x^{\alpha{\dot \beta}}, \vartheta^\alpha,
{\bar \vartheta}^{{\dot \alpha}})$,
the $N=1$ super Yang-Mills theory is conventionally described in terms of
two spinorial field strengths $w_\alpha, \bar w_{\dot \alpha} $ defined by
$$\eqalign{
[{\bar {\cal D}}_{{\dot \beta}} ~,~{\cal D}_{\alpha{\dot \alpha}}]
=&\ \epsilon_{{\dot \beta}{\dot \alpha}}\ w_\alpha       \cr
[{\cal D}_{\beta} ~,~{\cal D}_{\alpha{\dot \alpha}}]
=&\  \epsilon_{\beta\alpha} \bar w_{{\dot \alpha}}  ,\cr} \eqno(6) $$
where the gauge-covariant derivatives ${\cal D}_A \equiv \partial_A + A_A =
({\cal D}_{\alpha{\dot \beta}}, {\cal D}_\alpha,
{\bar {\cal D}}_{{\dot \beta}}) $ satisfy the familiar constraints
$$\{{\cal D}_{\alpha} ~,~{\cal D}_{\beta}\} =\ 0         \eqno(7a)$$
$$  \{{\bar {\cal D}}_{{\dot \alpha}} ~,~{\bar {\cal D}}_{{\dot \beta}}\}
=\ 0 \eqno(7b)$$
$$\{{\cal D}_{\alpha} ~,~\bar{\cal D}_{{\dot \beta}}\}
=\  2\ {\cal D}_{\alpha{\dot \beta}} \eqno(7c)$$
and the supertranslations $$\partial_A  =(\partial_{\alpha{\dot \beta}},
 D_\alpha, \bar D_{\dot \beta} ) \equiv (\der {x^{\alpha{\dot \beta}}},
 \der {\vartheta^\alpha} ,
 \der {{\bar \vartheta}^{{\dot \beta}}}
 + 2 \vartheta^\alpha \der{x^{\alpha{\dot \beta}}},) $$
realise the free superalgebra
$$[D_{\beta} ~,~\partial_{\alpha{\dot \beta}}]
=\  [\bar D_{{\dot \alpha}} ~,~\partial_{\alpha{\dot \beta}}] =\
\{D_{\alpha} ~,~D_{\beta}\}
=\ \{\bar D_{{\dot \alpha}} ~,~\bar D_{{\dot \beta}}\} = \ 0$$
$$[D_{\alpha} ~,~D_{{\dot \beta}}] =\  2\ \partial_{\alpha{\dot \beta}}. $$

The self-duality equations for the superconnection $A_A$
take the form of the following further (in comparison with (6))
constraints
$$[{\cal D}_{\beta} ~,~{\cal D}_{\alpha{\dot \alpha}}] =\ 0 , \eqno(8)$$
which say that $\bar w_{{\dot \alpha}}$ vanishes. That this is the
supersymmetrisation of the usual ($N=0$) self-duality condition
is evident from the dimension $(1 | 1)$ Jacobi identity which
yields the following superfield equations
$$ f_{{\dot \alpha}{\dot \beta}} =\ 0  \quad (a), \quad
{\cal D}_{\alpha} w_{\beta} =\ 2\ f_{\alpha\beta} \quad (b) \eqno(9)$$
for the self- and anti-dual
vector field strengths $f_{{\dot \alpha} {\dot \beta}}, f_{\alpha \beta}$
appearing in the definition
$$[{\cal D}_{\alpha{\dot \alpha}} ~,~{\cal D}_{\beta{\dot \beta}}]
\equiv\ \epsilon_{\alpha\beta} f_{{\dot \alpha}{\dot \beta}}  ~
+~\epsilon_{{\dot \alpha}{\dot \beta}} f_{\alpha\beta} .  $$
The dimension $({3\over 2} | 1)$ and $({1\over 2} | {3\over 2})$
Jacobi identities are then
satisfied, respectively, if $w_{\alpha}$ and $f_{\alpha\beta}$ satisfy the
(anti-) chirality equations
$${\cal D}_{\gamma} f_{\alpha\beta}=\ 0 \quad (c), \quad
\bar{\cal D}_{{\dot \beta}}w_{\alpha} =\ 0  \quad (d).   \eqno(9)$$
All other Jacobi identities are then automatic,
requiring the introduction of no further superfield strengths.
Equations (8) are the superfield self-duality equations. They
indeed imply the equations of motion
$ \epsilon^{\gamma\alpha}{\cal D}_{\gamma{\dot \gamma}}f_{\alpha\beta}
=\ \epsilon^{\gamma\alpha}{\cal D}_{\gamma{\dot \gamma}} w_\alpha =\ 0.$
We have therefore shown that the constraints (7), (8) for
$A_A$ imply the superfield equations (9) for the superfield-strength
$w_\alpha$ and superfield vector-potential $ A_{\alpha{\dot \beta}}$.
These in turn imply
the ordinary space  supersymmetric self-duality equations for
the component fields (which we denote by the same symbols as the
superfields of which they are the leading components)
$$f_{{\dot \alpha}{\dot \beta}} =\ 0, \quad
\epsilon^{\gamma\alpha}{\cal D}_{\gamma{\dot \gamma}} w_\alpha
=\ \bar w_{\dot \alpha} = \ 0  , $$
on eliminating all gauge degrees of freedom depending on the anticommuting
superspace coordinates. The converse, that given a set of component
fields satisfying these component equations, one can reconstruct superfields
satisfying (9), and in turn the superconnection $A_A$ satisfying (7), (8)
is also true. The proof closely follows the methods of [8].

{\bf 3.} Our main purpose here, however, is
to introduce yet another piece of data
corresponding to the above three: a `holomorphic' prepotential in harmonic
superspace. We shall show that the  constraints (7), (8)
imply generalised Cauchy-Riemann (CR) conditions for a prepotential
in harmonic superspace and that any superconnection satisfying (7), (8)
may be expressed
in terms of such holomorphic prepotentials. (We shall use `holomorphic' in
this generalised sense; to describe solutions of these generalised CR
conditions) . This construction
is a realisation of the twistor construction [3] for supersymmetric
self-dual systems in the harmonic superspace framework.

In the present complex setting, the harmonics $u^\pm_{\dot \alpha}$ remain,
as usual,
$S^2$ harmonics, the 2-sphere being a coset of the $SL(2,\C)_R$ part of
the Lorentz group with its maximal parabolic subgroup. In this setting,
$u^+$ and $u^-$ are {\it not} complex conjugates of each other; and are
defined up to parabolic subgroup transformations. They obey the usual
constraints $u^{+{\dot \alpha}} u^-_{\dot \alpha} =\ 1.$ Details of this
construction, the role of conformal invariance, as well as appropriate
reality conditions may be found in [6]. For our $N=1$ harmonic superspace
the derivatives
$  \partial^\pm_\alpha
 \equiv u^{\pm{\dot \alpha}} \partial_{\alpha {\dot \alpha}},\quad
\bar{D^+} = u^{\pm{\dot \alpha}}\bar D_{{\dot \alpha}} ,\quad D_\alpha ,$
together with harmonic ones
$$ D^{\pm\pm} =\ u^{\pm {\dot \alpha}} \der {u^{\mp {\dot \alpha}}},\quad
D^0 =\ u^{+{\dot \alpha}} \der {u^{+{\dot \alpha}}}
      - u^{-{\dot \alpha}} \der {u^{-{\dot \alpha}}},$$
realise the free superalgebra
$$\{D_{\alpha} ~,~\bar D^\pm \} =\  2\ \partial^\pm_{\alpha}, \eqno(10a)$$
$$ [D^{++} , D^{--} ] = D^0 ,\quad [D^0 , D^{\pm\pm}] =\ \pm 2 D^{\pm\pm}
\eqno(10b)$$
$$[D^{\pm\pm}, \bar D^\mp] = \bar D^\pm ,
\quad  [D^{\pm\pm}, \partial^\mp_\alpha] = \partial^\pm_\alpha, \eqno(10c)$$
with all other commutators vanishing.
Now, in terms of the gauge-covariant derivatives
$$ \nabla^+_{\alpha}  =    \partial^+_{\alpha} + A^+_\alpha
 =  u^{+{\dot \alpha}} {\cal D}_{\alpha {\dot \alpha}} ,\quad
  \bar{{\cal D}^+} =  \bar D^+  + \bar A^+
            =  u^{+{\dot \alpha}}\bar {\cal D}_{{\dot \alpha}}
,\quad {\cal D}_\alpha   =      D_\alpha                   ,$$
(recall that $A_\alpha = 0$) obeying the commutation relations
$$[D^{++}, \bar{\cal D}^+] = 0, \eqno(11a)$$
$$[D^{\pm\pm}, {\cal D}_\alpha] = 0,    \eqno(11b)$$
$$[D^{++}, \nabla^+_{\alpha}] = 0,  \eqno(11c)$$
the constraints (7), (8) are equivalent to the Cauchy-Riemann system
$$[\bar{\cal D}^+, \nabla^+_{\alpha}] = 0, \eqno(12a)$$
$$[{\cal D}_\beta, \nabla^+_\alpha] = 0,  \eqno(12b) $$
$$ [\bar{\cal D}^+, {\cal D}_\alpha] =\ 2\ \nabla^+_{\alpha} .\eqno(12c).
    $$
Remarkably, these are precisely the integrability conditions for equations
(1), (3) and (4).
We therefore have the following pure-gauge-like expressions for
$A^+_\alpha$ and $\bar A^+$
$$A^+_\alpha  =\   - \partial^+_\alpha \phi \phi^{-1}, \quad  \quad
\bar A^+ =\       -\bar D^+\phi \phi^{-1}. \eqno(13)     $$

{\bf 4.} Equations (11), (12) are therefore equivalent to the
constraints (7), (8). Let us now choose a coordinate basis, which we shall
call the analytic frame, in which the derivatives take the forms
$$\eqalign{
 \widehat {D_\alpha} =&\ \phi^{-1}[{\cal D}_\alpha]\phi
 = {\partial \over \partial\vartheta^\alpha}, \cr
 \widehat {\nabla^+_\alpha} =&\ \phi^{-1}[\nabla^+_\alpha] \phi =
 {\partial \over \partial x^{-\alpha}},\cr
 \widehat {\bar D^+} =&\  \phi^{-1}[\bar{\cal D}^+] \phi
  = {\partial \over\partial{\bar\vartheta^-}}
     + 2 \vartheta^\alpha {\partial \over \partial x^{-\alpha}},\cr
 \widehat {D^{++}} =&\  \phi^{-1}[D^{++}] \phi = D^{++} + V^{++} \cr
 \widehat {D^{--}} =&\  \phi^{-1} [D^{--}] \phi = D^{--} + V^{--}
.\cr} \eqno(14)$$
In this basis the covariant derivatives
$\widehat{\bar D^+}$ and
the $\widehat{\nabla^+_\alpha}$ become flat, losing their connections
($\widehat {\cal D}_\alpha$ remains flat), while
the harmonic derivatives $D^{\pm\pm}$ clearly acquire the connections
$$V^{++}  = \phi^{-1} D^{++} \phi ,\eqno(15a)$$
$$V^{--}  = \phi^{-1} D^{--} \phi .\eqno(15b) $$
In order to preserve the operator $D^0$ as a charge
counting operator, we have used (as usual, see e.g. [3]) the
conventional gauge in which it does not acquire a connection:
$$[\widehat D^{++}, \widehat D^{--}] =\ D^0.\eqno(16)$$
In this basis the dynamical content is contained entirely in (11),
the rest of the equations being kinematical.
Using the identity
$$ A ( B f f^{-1}) \equiv f ( B (f^{-1} A f)) f^{-1} + [A,B]f f^{-1},$$
for arbitrary differential operators $A$ and $B$,
eqs. (11) take the form of generalised CR conditions
$$\eqalign{
 {\partial \over\partial {\bar\vartheta^-}} V^{++} = 0, \cr
 {\partial \over\partial {\vartheta^\alpha}} V^{\pm\pm} = 0, \cr
 {\partial \over\partial {x^{-\alpha}}} V^{++} = 0. \cr}\eqno(17) $$
$V^{++}$ is therefore holomorphic: it depends on $x^{+\alpha}$,
 $\vartheta^+$ and $u^\pm$, being independent of $x^{-\alpha},
 \bar\vartheta^- $ and $\vartheta^\alpha$;
both $V^{++}$ and $V^{--}$ are chiral. Note that the third equation
is a consequence of the other two.
We have shown that to any solution of the super self-duality
constraints (8), there corresponds a chiral holomorphic superfield
$V^{++}$ taking values in the gauge algebra and
having component expansion
$$ V^{++}(x^{+\alpha}, \bar\vartheta^+,u^\pm_{\dot \alpha})
= v^{++}(x^{+\alpha}, u^\pm_{\dot \alpha})
+ \bar \vartheta^+ \chi^+(x^{+\alpha}, u^\pm_{\dot \alpha}).\eqno(18)$$
The superfield $V^{++}$ is defined modulo gauge transformations
$$ \delta V^{++} =\  [\widehat D^{++}, \lambda ]          ,$$
where $\lambda$ is an arbitrary holomorphic superfield.
Note that due to the presence of the fermion mode there is
an important difference with the $N=0$ case: whereas
the $N=0$ connection was encoded in one function on the 2-sphere,
in the present case we have {\bf two} functions, $v^{++}$ and $\chi^+$,
instead.

{\bf 5.} The converse statement, that any chiral analytic superfield
prepotential
$V^{++}$ encodes a superconnection $A_A$ satisfying (7), (8) also holds.
To reconstruct the superconnection $A_A$ there are two options.
\item{a)} We can start with the chiral superfield  $\phi$.
In this case
we need to solve (15a) for $\phi$; $V^{++}$ being given. Equations on
the 2-sphere of this kind are not too easy to solve and they appear in many
applications of the harmonic-twistor approach, see, e.g. [4].
In order to determine the corresponding superconnection solving (7), (8),
we need to insert the $\phi$ thus obtained into the following formulae
$$A_{\alpha{\dot \alpha}}
= 2\int d^2u ~u^-_{{\dot \alpha}} \phi \partial^+_\alpha  \phi^{-1} ~,
\quad \quad A_{{\dot \alpha}}
= 2\int d^2u ~u^-_{{\dot \alpha}}  \phi \bar D^+  \phi^{-1}, \eqno(19)$$
which follow immediately from (13); and $A_\alpha$ is of course zero.

\item{b)} Instead of $\phi$, we can start with the harmonic connection
$V^{--}$. It follows from (16) that
$$ Z \equiv  D^{++}V^{--} - D^{--}V^{++} + [V^{++} , V^{--}] = 0.
\eqno(20) $$
For a given holomorphic $V^{++}$ taking values in the gauge algebra,
it contains a set of coupled first-order linear equations for
the  gauge algebra components of $V^{--}$. These may be solved [9]
somewhat more easily than (15a). Now, as further consequences
of gauge-covariantising the harmonic derivatives (14), we have, from (10),
$$ [\widehat D^{--}, \bar D^+] =\  {\bar {\cal D}}^- ,\quad
  [\widehat D^{--}, {\bar {\cal D}}^-] =\  0 , \quad
  [\widehat D^{--}, \partial^+_\alpha] =\  \nabla^-_\alpha ,\quad
  [\widehat D^{--}, \nabla^-_\alpha] =\  0, \eqno(21)$$
from which we obtain superconnections
$$ \bar A^- = - \bar D^+ V^{--} ,\quad
A^-_\beta = - \partial^+_\beta  V^{--},\eqno(22)$$
in terms of solutions of (20). Now, superconnections satisfying (7),(8)
may be recovered by harmonic integration similar to (19).
In fact, from $V^{--}$  we may also directly construct
the superfield strength $w_\alpha$  satisfying
the superfield equations (9); namely, from (6),
$$ w_\alpha =\ -[\bar{\cal D}^+, \nabla^-_\alpha]
      =\ \bar D^+ \partial^+_\alpha V^{--}.\eqno(23)   $$

An alternative to equation (20) follows
from the following commutators contained in the superalgebra (10) (in the
analytic frame)
$$ [\bar {\cal D}^- , \nabla^-_\alpha] =\ 0
          = [\nabla^-_\alpha, \nabla^-_\beta] .$$
These yield the alternative equations for $V^{--}$ :
$$L^{--}_\alpha \equiv\
 -\partial^+_\alpha \bar D^- V^{--} + \partial^-_\alpha \bar D^+ V^{--}
+ [\bar D^+ V^{--}, \partial^+_\alpha V^{--}] =\ 0,\eqno(24) $$
$$ L^{--} \equiv\
 \partial^{+\alpha} \partial^-_\alpha V^{--}
 + [\partial^{+\alpha} V^{--}, \partial^+_\alpha V^{--}] =\ 0.
\eqno(25)$$
The latter equation is in fact the one introduced for the
$N = 0$ case in [10]. We note the following interesting interrelations
amongst the left-hand-sides of equations (20), (24) and (25):
$$\bar D^+ L^{--} =\
\partial^{+\alpha} L^{--}_\alpha, \quad \nabla^-_\alpha \partial^{+a} Z =\
\widehat D^{++} L^{--}, \quad \nabla^-_\alpha \bar D^+ Z =
\widehat D^{++} L^{--}_\alpha .$$

{\bf 6.} We now present an action for super self-duality.
Since all we need to have is the generalised CR conditions (17), with
$V^{++}$ expressed in terms of the field $\phi$, as a variational
equation, we plug this condition into an action functional with
the help of a
Lagrange multiplier-type auxiliary field. The latter does not propagate if
it contains only gauge degrees of freedom.
For the chiral superfield $\phi$, the action functional
$$ S =\  \int d^4x~d^2{\bar \vartheta}~du~~ tr ~
{}~(\bar D^+\zeta^{-3}  \phi^{-1} D^{++} \phi) \eqno(26)$$
yields, on varying the auxiliary field $\zeta^{-3}$, the CR condition
$\bar D^+ V^{++} = 0$, $V^{++}$ is chiral by definition; and the
final condition $\der{x^{-\alpha}} V^{++} = 0$ is a consequence.
Now, on varying $\phi$, we obtain
$$ -\phi^{-1} D^{++}[\phi \bar D^+ \zeta^{-3} \phi^{-1}] = \ 0.
\eqno(27a)$$
It follows (cf. [11]) that
$$ \bar D^+ \zeta^{-3} = 0.\eqno(27b)$$
All solutions of this equation have the form
$$\zeta^{-3} = \bar D^+ y^{-4}. \eqno(28)$$
However $\zeta^{-3}$ enters the action via $\bar D^+\zeta^{-3}$, so it
is only defined modulo the addition of $\bar D^+ y^{-4}$.
$\zeta^{-3} $ therefore does not represent any additional physical degree
of freedom. For $N = 0$ an analogous action was discussed in [11,12].

Alternatively, we may choose  $V^{--}$ (instead of $\phi$) as the
dynamical field, express
$V^{++}$ in terms  $V^{--}$ with the help of (20), and construct
an action having analyticity conditions for the functional $V^{++}[V^{--}]$
as variational equations.

There also exists the possibility (analogous to the $N = 0$ action
considered in [10]) of writing an action for eq.(25)
trilinear in \ $V^{--}$. It explicitly
contains a constant harmonic factor of charge $+4$, say $(u^+_1 u^+_2)^2$,
and is consequently not Lorentz invariant.

{\bf 7.} To conclude we generalise our construction to N-extended
harmonic-superspace with coordinates
$ x^{\pm_\alpha} ,\  \vartheta^{\alpha i} ,\  {\bar \vartheta}^\pm_j
\equiv u^\pm_{\dot \alpha} {\bar \vartheta}^{{\dot \alpha}}_j ,\
u^\pm_{\dot \alpha} $, where $i,j = 1,...,N$.
Solutions of the generalised CR conditions
$$\eqalign{
{\partial \over\partial {\bar\vartheta^-_j}}  (\phi^{-1} D^{++}\phi )  =& 0,
\cr
{\partial \over\partial {\vartheta^{\alpha i}}} (\phi^{-1} D^{\pm\pm}\phi )
=& 0, \cr
{\partial \over\partial {x^{-\alpha}}} (\phi^{-1} D^{++}\phi )   =&  0  \cr}
\eqno(29)$$
encode N-extended self-dual superconnections.
It may be verified that the superconnection components
$$ A_{\alpha{\dot \alpha}}
= \int d^2u ~u^-_{{\dot \alpha}} \phi \partial^+_\alpha  \phi^{-1} , $$
$$ A^j_{{\dot \alpha}}
= \int d^2u ~u^-_{{\dot \alpha}}  \phi \bar D^{+j}  \phi^{-1}       ,$$
$$   A_{\alpha i} =  0  , $$
with $\phi$ satisfying (29), automatically satisfy the
self-dual restrictions of the conventional extended superconnection
constraints
$$ F_{i\alpha j\beta} = 0
= F^{ij}_{{\dot \alpha}{\dot \beta}} + F^{ij}_{{\dot \beta}{\dot \alpha}} $$
$$F^{i}_{\alpha{\dot \beta} j} =0 .$$
That integrability conditions for these equations yield (29) follows from
reasoning parallel to that for the N=1 case above.

For the full (non-self-dual) N = 2 and 3 theories, for which harmonic
superspace
formulations (harmonising the internal automorphism group) exist [4,6],
the self-duality conditions are also equivalent to ``double''
analyticity conditions arising from Lorentz as well as internal automorphism
group harmonisation. We intend to return to these theories elsewhere.

{\bf Acknowledgements}

We should like to sincerely thank
A. Galperin, A. Leznov, E. Ivanov, D. Khetselius and E. Sokatchev
for valuable discussions. One of us, VO, appreciates very much the
fruitful and cordial atmosphere of the CERN Theory Division.
\vskip 1.5 true cm

{\bf References}

\item{[1]} R.S.~Ward, Phil.Trans.Roy.Soc. A315 (1985) 451; N.J.~Hitchin,
Proc.Lond.Math.Soc. 55 (1987) 59.
\item{[2]} V.~Novikov, M.~Shifman, A.~Vainstein and V.~Zakharov,
Nucl.Phys. B229, (1983), 394, 407;  A.~Semikhatov, Phys.Lett. 120B
(1983) 171; I.~Volovich, Phys.Lett. 123B (1983) 329; S.~Ketov,
H.~Nishino, S.J.~Gates, Jr, prepr. UMDEPP 92-211.
\item{[3]} For instance, R.S. Ward and R.O. Wells, Twistor Geometry and
Field Theory, (CUP, Cambridge, 1990).
\item{[4]} A. Galperin, E. Ivanov, S. Kalitzin, V. Ogievetsky and
E. Sokatchev, in Quantum Field
Theory and Statistics,(A.Hilger, Bristol, 1987) vol.2, 233;
O. Ogievetsky, in Proc. Conf. on Group. Theor. Methods in Physics, Varna,
1987 (Berlin, Springer, 1988) 548;
Thesis, Lebedev Physical Institute (1988) 329, Moscow;
S. Kalitzin and E. Sokatchev, Class. Quantum Gravity 4 (1987) L173.
\item{[5]} A. Galperin, E. Ivanov, V. Ogievetsky and E. Sokatchev,
Ann. Phys. (N.Y.) 185 (1988) 1.
\item{[6]} M. Evans, F. G\"ursey and V. Ogievetsky, prepr.
CERN-TH 6533/92, RU-92-11-B.
\item{[7]} F, Delduc, A. Galperin and E.Sokatchev, Nucl. Phys. B368,
(1992) 143.
\item{[8]} J. Harnad, J. Hurtubise, M. L\'egar\'e and S. Shnider,
Nucl. Phys. B256 (1985) 609; J. Harnad and S. Shnider, Commun. Math.
Phys. 106 (1986) 183.
\item{[9]} B.Zupnik, Theor.Math.Phys. 69 (1986) 175.
\item{[10]} A. Leznov, Theor.Math.Phys. 73 (1987) 302;
A. Leznov and M. Mukhtarov, J. Math. Phys. 28 (1987) 2574.
\item{[11]} S. Kalitzin and E. Sokatchev, Phys.Lett. 257B (1991) 151.
\item{[12]} N. Markus, Y. Oz and S. Yankielovicz, Nucl. Phys. B379 (1992)
121.
\end